\documentclass[prd,preprint,aps]{revtex4-2}

\usepackage{graphicx}
\usepackage{amsmath}
\usepackage{amssymb}

\usepackage{comment}
\usepackage{subfig}
\usepackage{caption}
\usepackage{color}
\usepackage{url}
\usepackage{hyperref}

\begin{document}

\title{Axion Gamma-Ray signatures from Quark Matter in Neutron Stars and Gravitational Wave Comparisons}
\author{Bijan Berenji}

\affiliation{fiziSim LLC \\  2245 Hillsbury Rd. \\Westlake Village, CA 91361 \\ bijan.berenji@fizisim.com }
\affiliation{SLAC National Accelerator Laboratory \\ 2575 Sand Hill Rd., Menlo Park CA 94025 \\ bijanb@slac.stanford.edu}
\maketitle

\date{\today}
		\section{Abstract}
Axions are hypothetical particles proposed to solve the strong CP problem in QCD and may constitute a significant fraction of the dark matter in the Universe.  Axions are expected to be produced in neutron stars and subsequently decay, producing gamma-rays detectable by the Fermi Large Area Telescope (Fermi-LAT).   Considering that light QCD axions, as opposed to axions $>1$eV, may travel a long range before they decay into gamma rays, neutron stars may appear as a spatially-extended source of gamma rays.  We extend our previous search for gamma rays from axions, based on a point source model,  to consider the  neutron star as an extended source of gamma rays.  The extended consideration of neutron stars' leads to higher sensitivity to searches for axions, as it will be shown.   
	We investigate the spatial emission of gamma rays using phenomenological models of neutron star axion emission.    We present models including the fundamental astrophysics and relativistic, extended gamma-ray emission from axions around neutron stars.  A Monte Carlo simulation of the LAT gives us an expectation for the extended angular profile and spectrum.   For a source of $\simeq$ 100 pc, we predict a mean angular spread of $\simeq 2^\circ$ with gamma-ray energies in the range 10-200 MeV, due to the cutoff of the spin-structure function $S_{\sigma}(\omega)$.   We demonstrate the feasibility of setting more stringent limits for axions in this mass range, excluding a range not probed by observations before.  We consider projected sensitivities for mass limits on axions from J0108-1431, a neutron star at a distance of 130 pc.  Based on the extended angular profile of the source, the expected sensitivity of the 95\% CL upper limit on the axion mass from J0108-1431 is  $\lesssim$10 meV.    The limit based on 7.9 years of Fermi-LAT data is 0.76 meV for an inner temperature of the neutron star of 20 MeV.  In this work, we consider only QCD axions, but the results may of course be generalized to axion-like particles (ALPs). 

\section{Introduction}

The QCD axion is being investigated for compelling theoretical reasons, and many promising methods have been investigated for its detection.   The axion was postulated to solve the strong $CP$ problem~\cite{PQ}, by the mechanism of spontaneously broken U(1) PQ symmetry,  and may constitute a significant fraction of the dark matter in the universe~\cite{Sikivie201122}.   Axions can be studied by means of neutron stars, from  which they are theorized to be produced by nucleon-nucleon bremsstrahlung.  The coupling of axions to the electromagnetic field can also generate axions by the Primakoff effect~\cite{Skobolev2000}.

According to our previous model of axion emission from neutron stars, gamma rays may be produced from axion decays from neutron stars, creating a point source of gamma rays~\cite{berenjiAxions}.   In this paper, we extend the model of emission of axions to consider the neutron star as an extended source of gamma rays.  Considering that light axions may travel a long range before they decay into gamma rays, neutron stars may in fact be a spatially-extended source of gamma rays.  The limits on the axion mass, which were placed from Fermi Large Area Telescope (LAT) analysis of the point source model on neutron star targets, might be improved with consideration of the spatial extension of the gamma rays, as it increases the solid angle the gamma-rays subtend around the neutron star.

Observations with the Fermi LAT are crucial to an axion search or setting limits on axion parameters.  Here, for the first time, we use Fermi-LAT observations of  neutron stars with an extended source model to search for signatures of axions. The Fermi LAT is an imaging, wide field-of-view, pair-conversion telescope which detects gamma rays with energies from 20 MeV to over 300 GeV~\cite{instrumentPaper}.  This energy range includes the energies of photons from decaying axions, roughly 30 to 200 MeV in our model.

	Extended gamma-ray sources have been extensively studied with the Fermi LAT, e.g., Ref.~\cite{lande2012search}, including pulsar wind nebulae and supernova remnants.   A search for extended sources in the galactic plane, detecting 46 sources, has also been performed with the Fermi LAT~\cite{ext2017}.  In addition, a search for extended high latitude sources has found 24 sources that demonstrate extension~\cite{extHiLat}. Further, dark matter in galaxies may be modeled as extended sources of gamma rays, e.g., Ref.~\cite{AbazajianDM}. In addition, Andromeda has recently been observed as an extended source~\cite{andromedaExtended}. We may note that spatially-extended emission from axions may occur in the vicinity of supernova remnants.  Decays that occur at a distance from supernova remnants have been considered in Ref.~\cite{giannotti}.   In addition, there have been many investigations of photon-axion and photon-ALP conversion in extragalactic magnetic fields, with large distance scales, e.g., Ref.~\cite{meyer2013first}.  We consider the decays, but do not consider the oscillations, because the distances are not large~\cite{raffeltConversion}. Here, we consider variation on the point-source neutron star model considered previously, and consider extended emission due to axions decaying at a certain distance away from the source.  As the distribution of gamma rays arising from axion decays falls off not as rapidly as a point source, according to this theory, we model the distribution of axions as a spatially extended source.  According to the theory of convolution, the convolution of a flux with a delta-function gives back the original flux function;  however, the convolution of a flux with a distribution with a width of 2-3$^\circ$ is noticeably different, even if the PSF is larger than the width.   

	If there is no signal detected, the limits may potentially be improved with respect to the point source analysis of neutron stars, due to the photons from axion decay potentially being spread out over a larger solid angle in the regions of interest (ROIs) corresponding to the neutron stars we wish to investigate. 

There are theoretically and observationally justifiable reasons for investigating this model. The theoretical lower bound on axion decays from supernova energy-loss arguments has been placed at $\sim 10$ meV~\cite{raffelt}.  The 100 meV to 1 meV range has been mentioned as a promising region for future axion searches~\cite{raffelt2011meV, redondo2012journey}.  The possibility of diffuse emission from axions produced by $NN-$bremsstrahlung in supernova cores has been theorized to yield axion mass limits in the meV range~\cite{raffelt2011meV}.  We provide a model that leads to  more restrictive constraints on the axion mass from considering extended axion emission from neutron stars.  There are recent theoretical constraints from neutron star cooling which predict axion masses in a close mass range~\cite{sedrakian}.  It is possible to detect gamma-rays in the energy range 30-200 MeV with the Fermi LAT~\cite{pass8}. Furthermore, the model projects a flux from 10 meV axions that can be measured by Fermi LAT, as it will be shown.   
 
		Several investigators have recently studied axions via nucleon coupling, such as with axion decay, with neutron star cooling, and solar axions.  In Ref.~\cite{lloyd}, a sample of 17 isolated neutron stars were investigated with similar spectral models for axion decay as determined by Ref.~\cite{berenjiAxions}, to determine a 95\% C.L. upper limit on $m_a<9.6\cdot 10^{_3}$eV.  From Ref.~\cite{hamaguchi}, an upper limit of 0.01 eV was placed from neutron star cooling of the neutron star at the center of the supernova remnant Cas A, relying upon a model of neutron superfluidity.     In Ref.~\cite{abdelhameed}, the 8.41 keV line for $^{169}$Tm resonant excitation was studied for solar axions produced via the Primakoff effect, yielding a 90\%C.L. upper limit of 24 eV.  

The relation between $m_a$ and the axion coupling $f_a$ is given by
	\begin{equation} m_a = \frac{0.6 {\rm meV}}{f_a/(10^{10}\ {\rm GeV})}\label{mafa} \end{equation}

Our search for axions from neutron stars depends on the axion-coupling to quarks via $NN$-bremsstrahlung, where the derivative couples to the axion field in the Lagrangian as: 
\begin{equation} \mathcal{L} \subset \frac{1}{f_a} g_{ann}\left(\partial_\mu a\right) \bar{N} \gamma^\mu \gamma_5 N. \end{equation}
	
The axion-nucleon coupling may be parametrized in terms of $m_a$: 
\begin{equation} g_{ann} = 10^{-8}\left(\frac{m_a}{1\ {\rm eV}}\right). \end{equation} 
	We consider KSVZ axions~\cite{ksvz1,ksvz2}, to be distinguished from the DFSZ axion model~\cite{dfsz1,dfsz2}.  According to the KSVZ ``hadronic'' axion model, the heavy quarks are electrically neutral and carry PQ charges.  On the other hand, in the DFSZ model, there are at least two Higgs doublets and ordinary quarks have PQ charges~\cite{RPP}.  The axion field should be a Bose-Einstein Condensate (BEC)~\cite{nambu}, and is expected to be responsible for the nucleon electron dipole moment(EDM).

	In this article, we investigate the spatial emission of gamma rays using phenomenological models in order to determine the projected sensitivities of Fermi-LAT observations from photon fluxes of neutron stars.  We present the fundamental astrophysical model, the model for extended gamma-ray emission from axions around neutron stars, the Monte Carlo simulation model.  We demonstrate the feasibility of setting more stringent limits for QCD axions than previous literature values~\cite{sanchez2009hints}, which could potentially exclude a range not probed by observations before.  

	In Section~\ref{sec:astroPhModel}, we present the underlying astrophysical model. In Section~\ref{sec:extended}, we discuss the extended emission of gamma rays due to axions from neutron stars. In Section~\ref{sec:proj}, we discuss projected limits from neutron star J0108-1431.   In Section~\ref{sec:obs}, we present the methods of and results for analysis of Fermi LAT observations of the neutron star J0108-1431.  In Section~\ref{sec:disc}, we discuss the relevance of the limits in the astrophysical context as well as to other astrophysical limits on the axion mass.

\section{Astrophysical Model}\label{sec:astroPhModel}

\begin{figure} 
\begin{centering}
\includegraphics[width=6in]{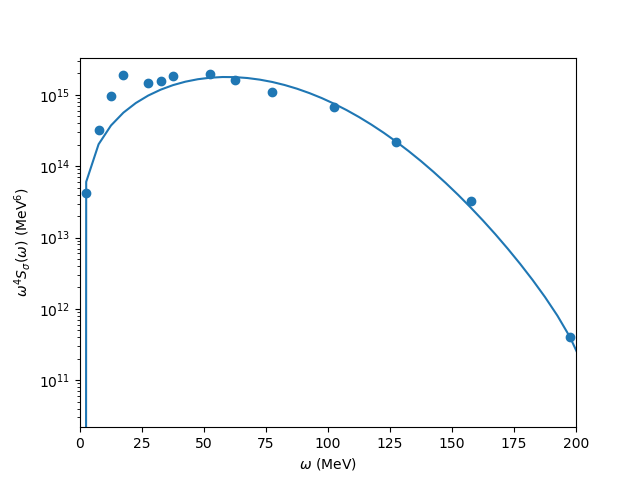}
\caption{The function $g(\omega)=\omega^4 S_\sigma(\omega)$, whose parameters have been fit to an analytic functional form given by Equation~\ref{eq:homega}.  Dots refer to Monte Carlo simulated points, while the solid line refers to fitted model described by Equation~\ref{eq:homega}.\label{fig:omegaFcn}}
\end{centering}
\end{figure}

Axions may be produced in neutron stars by the $NN$-bremsstrahlung reaction $n n \to n n a$, where $n$ is a neutron~\cite{SNbounds}. The axions produced in this manner would be relativistic (see below). For a physical description of this process, we follow the phenomenology of Hanhart, Philips, and Reddy~\cite{uw}, who model this process as a nucleon-nucleon scattering process.
We developed an astrophysical model to derive an energy flux from axions emitted from neutron stars, which subsequently decay to photons in Ref~\cite{berenjiAxions}.  In deriving the differential photon flux ($\Phi$), we consider the differential emissivity with respect to axion energy.  In the case of radiative decay of axions $a\to 2\gamma$, we determine the photon energy from the axion mass, the relativistic boost $\gamma$, and the angle of photon direction with respect to the axion direction, $\theta_a$.  In addition, we consider a neutron star of volume $V_{NS}$ as a uniform density sphere with a radius of 10 km, a timescale for axion emission $\Delta t$ as described below, a neutron star at a distance $d$, and the axion decay width $\Gamma_{a\gamma\gamma}$.   We consider $\Gamma_{a\gamma\gamma}$ as given by~\cite{axionBook}:

\begin{equation} \Gamma_{a\gamma\gamma}=1.1\times 10^{-24} \ {\rm s}^{-1} \left(\frac{m_a}{1 {\rm eV}}\right)^5. \end{equation}
 The energy flux is related to the axion emissivity from nucleon-nucleon bremsstrahlung, as well as the timescale of axion emission from the nuclear medium, which both depend on the axion mass:
\begin{equation} E\frac{d\Phi}{dE} = 2 \frac{d\epsilon_a}{d\omega} \delta(E-\omega/2) \frac{V_{NS} \Delta t \Gamma_{a\gamma\gamma} }{4\pi d^2}, \label{eq:energyFlux}\end{equation}
$\omega$ is the energy of emitted axions, $\epsilon_a$ is the axion emissivity of the neutron star matter. This phenomenologically accounts for the nucleon-nucleon bremsstrahlung process as a nucleon-nucleon scattering.   $V_{NS}$ is the volume of the neutron star, and $d$ is the distance to it.  We model the timescale of axion emission from a neutron star as the mean time $\Delta t$ between successive axion emissions in the nuclear medium.  
It is shown in Ref.~\cite{berenjiAxions} that $\Delta t \simeq 23.2$ s.  
\noindent Upon simplification of equation~\eqref{eq:energyFlux}, we obtain the following equation for the spectral energy distribution:
\begin{equation} E \frac{d\Phi}{dE} = 1.8 \times 10^{-2} \left(\frac{m_a}{\rm eV}\right)^5 \left(\frac{\Delta t}{23.2 \ {\rm s}}\right)\left(\frac{100 \ {\rm pc}}{d}\right)^2  \left(\frac{2E}{100 \ {\rm MeV}}\right)^4 \left(\frac{S_\sigma(2E)}{10^7 \ {\rm MeV}^2}\right) \ {\rm  cm}^{-2} {\rm s}^{-1}. \label{eq:sed}\end{equation}
 where $S_\sigma(\omega)$ is the spin-structure function, which accounts for the energy and momentum transfer and includes the spins of the nucleons. 
The function $g(\omega)=\omega^4 S_\sigma(\omega)$, which we use in our extended analysis, is shown in Figure~\ref{fig:omegaFcn}.  For the purpose of this investigation, this function has been fit to an analytic functional form, according to a log-likelihood minimization using the MINUIT optimizer, given by

\begin{equation} g(\omega) = \left(\alpha+\beta\omega\right)\gamma\exp\left(-\frac{1}{2}\left( (\omega-\delta)/\epsilon \right)^2\right)\exp\left( \eta + \theta\omega\right)\label{eq:homega}
 \end{equation} 
	In Table~\ref{tab:param}, we present the values of the parameters of $g(\omega)$.  

 \begin{table}
	 \begin{centering}
	 \begin{tabular}{|l|l|}
		 \hline
		 parameter & best-fit value \\
		 \hline
	 $\alpha$ & $-1.29\times 10^{11}$ \\
	 $\beta$ & $-0.364$ \\
	 $\gamma$ & $-1.37\times 10^{4}$ \\
	 $\delta$ & $31.3$ \\
	 $\epsilon$ & $37.0$ \\
	$\eta$ & $22.6$ \\
	$\theta$ & $3.49\times 10^{-3}$ \\
	 \hline
 \end{tabular}
	 \caption{Best-fit parameters of the function $g(\omega)$ as parametrized by equation~\eqref{eq:homega}}.\label{tab:param}
	 \end{centering}
	 \end{table}

We may consider the effect on axion mass limits due to variations in the model parameters.  
In the model of neutron stars that we are considering~\cite{ruster2005phase}, we may consider   $T=20$ MeV, and $\mu/T=10$~\cite{ruster2005phase}.  In the model, the neutron star matter has superfluidity, and its phase diagram is described by QCD.   Since the observed surface temperature of J0108-1431 is 0.28 keV~\cite{pavlov2009}, our assumption of the core temperature of $T=20$ MeV with a modest temperature gradient can be justified. The source J0108-1431 is an isolated neutron star which is detected in X-rays but not in $\gamma$-rays~\cite{garmire}.   Furthermore, superfluidity has been demonstrated to reduce the late time cooling of neutron stars, up to an age of $10^9$ yr., by the mechanism of frictional heating, as well as suppressing the neutrino emission energy loss mechanism~\cite{riper}.  
\section{Extended Emission of Axions from Neutron Stars}\label{sec:extended}

Axions decay with finite width $\Gamma_{a\gamma\gamma}=1.1\times 10^{-24} \ {\rm s}^{-1} \left(\frac{m_a}{1 {\rm eV}}\right)^5$; the probability that axions decay from the point of emission from the neutron star increases with distance from the source.  In other words, the survival probability decreases with angular distance from the neutron star.  Thus, the gamma rays arising from the axion decay would render the neutron star as an extended source in gamma rays.  The differential survival probability $dP$ is related to the probability $P$ as follows:

\begin{equation} dP = -P \frac{\Gamma_{a\gamma\gamma}}{\beta \gamma c} dr. \end{equation}
   
In the preceding equation, we divide $dr$ by $\beta c$, to obtain the time to traverse a radial distance $dr$.  
We also divide the decay rate $\Gamma_{a\gamma\gamma}$ by $\gamma$ and $\beta$, the commonly-used relativistic parameters, to account for time dilation. 
\begin{equation} P_0 = \frac{\Gamma_{a\gamma\gamma}}{\beta\gamma c} \end{equation}

The energy conservation condition, that the sum of kinetic energy plus potential energy of a radiated axion be equal to the energy radiated by the axion from the neutron star, can be expressed as:
\begin{equation}  K + U = -E_{rad} \end{equation} 
The preceding equation leads us to the following expression in terms of energy densities:
\begin{equation}  \frac{1}{2}\rho\beta^2 - \frac{G\rho M_{NS}}{c^2 r} = -\frac{\int \ dt \int d\Omega \int_0^{R_{NS}} dr \ r^2  \epsilon_a }{4/3 \pi r^3 c^2}, \end{equation}
where $\rho$ is the mass density of axions, $M_{NS}$ is the mass of the neutron star, and $G$ is the gravitational constant. 
In the limit $G M_{NS}/(rc^2) \gg \beta^2$, we obtain a distribution of axions $\rho\sim 1/r^2$. This is assuming a sufficient emissivity of axions from the neutron stars.  
Thus, we may convolve the function $P(r)$ with the function $1/r^2$, which describes the spatial density distribution of axions, to obtain $f(r;\gamma)$ .  This is justified because the probability of being found at a distance $r$ ($\sim 1/r^2$) is mutually exclusive of the probability of survival at a distance $r$, i.e., the joint probability distribution is the convolution of these two functions:
 

The energy and spatial dependence of the flux may be factorized: 

\begin{equation} \frac{d\Phi}{dEd\Omega} = \frac{1}{2\pi} \frac{dP_r(\theta;m_a,\omega)}{d\cos\theta} \frac{d\Phi_0}{dE}.   \end{equation}

\begin{figure}
\begin{centering}
\includegraphics[width=4in]{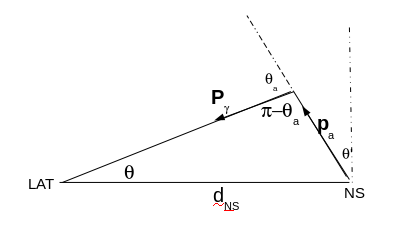}
\caption{The geometry of the axion decay into photon. The NS-LAT line defines the focal plane.  The axion is emitted on a radial path with a colatitude of $\theta^\prime$.  The decay photon is emitted at an angle $\theta_a$, and the $\theta$ is a measure of the extension of the source.\label{fig:geom}}
\end{centering}
\end{figure}

The geometry of the decays with respect to the neutron star and the LAT is shown in Figure~\ref{fig:geom}.
The distribution of the opening angles of the photons, $\theta_a$, with respect to the axion momentum direction, is given by~\cite{pionDecay}:

\begin{equation} P(\theta_a) = \frac{1}{4\beta \gamma} \frac{\cos\frac{\theta_a}{2}}{\sin^2\frac{\theta_a}{2}} \frac{1}{\sqrt{\gamma^2\sin^2\frac{\theta_a}{2}-1}}. \end{equation}

This distribution is strongly peaked in the forward ($\theta_a=0$) direction, and has a characteristic width $~1/\gamma$.  We may determine the spatial distributions of $\gamma$-rays using the following procedure.   We sample a $\gamma$ parameter from the distribution $g(\omega) = \omega^4 S_\sigma(\omega^4)$.
The radial coordinate of decay is sampled randomly from $f(r;\gamma)$.  We can determine $\theta$ simply from geometrical considerations according to

\begin{equation} \theta = \arcsin\left(\frac{r}{d}\sin(\pi - \theta_a)\right). \end{equation}

Geometrically, the condition for acceptance of the photon event is:

\begin{equation} \left|\theta - (\pi/2 - \theta_a - \theta_p) \right| < 0.4 \ {\rm rad} \end{equation}
This corresponds to a condition on the polar angle $\theta_p \lesssim 23.0^\circ$, where $\theta_p$ is the angle between the normal vector $\hat{n}$ to the top of the LAT and the vector of the photon momentum $\hat{p}_\gamma$.  at any given instant, 
This condition on the polar angle is derived as follows:
define a colatitude angle to $\theta^\prime$, $\theta_1$.
From Figure ~\ref{fig:geom}, it follows that 
\begin{equation} \theta_1 = \theta_a - \theta, \end{equation}
	and
	\begin{equation} \theta^\prime = \pi/2 - \theta_a + \theta, \end{equation}

		Considering that the field of view (FOV:=$\Delta\Omega_{\rm LAT}$) of the LAT is 2.4 sr~\cite{instrumentPaper}, where
		\begin{equation} \Delta \Omega_{\rm LAT} = \int_{\rm LAT} d\phi \int_{\rm LAT} d(\cos\theta)  \end{equation}

			\begin{equation} \Delta \Omega_{\rm LAT} = 2.4 = 2\pi \int d(\cos\theta) \end{equation}
			Thus, we derive the condition of the angular acceptance of the photon events in the LAT as:
				
				\begin{equation} \Delta \cos\theta \simeq 0.4. \end{equation}

\begin{figure}
\begin{centering}
	\includegraphics[width=6cm]{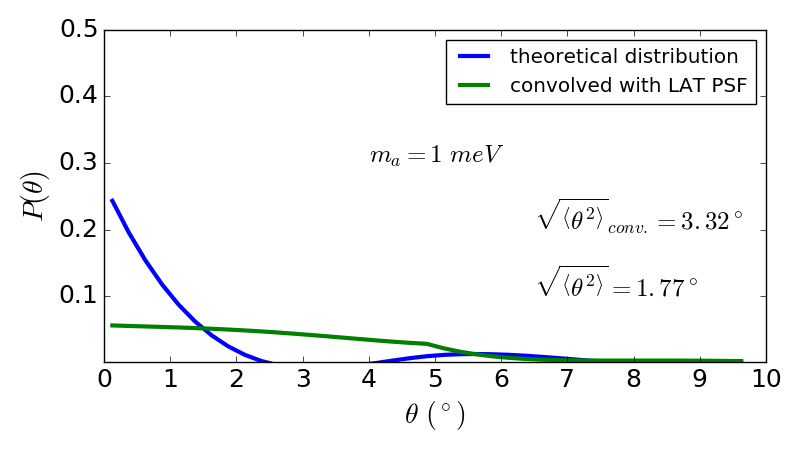}
	\includegraphics[width=6cm]{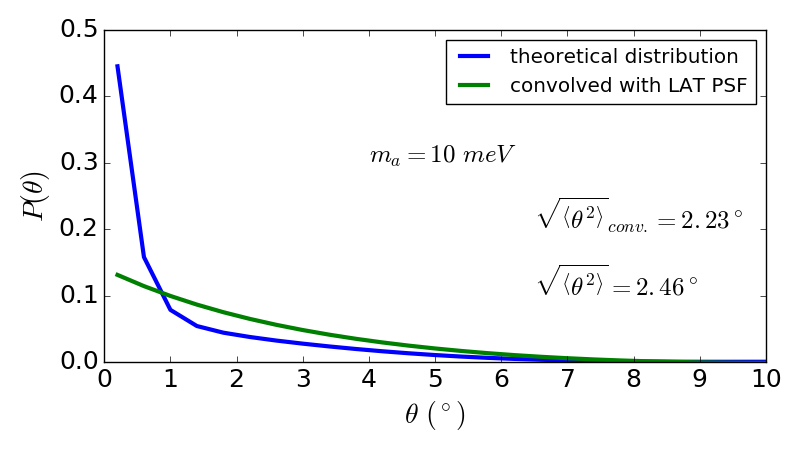}
	\caption{Angular profile of 1 meV and 10 meV axions: theoretical distribution (blue) and theoretical distribution convolved with the Pass8 LAT PSF at 60 MeV (green).}\label{fig:th}
\end{centering}
\end{figure}

In Figure~\ref{fig:th}, we plot the angular distributions for $P(\theta)$ and $P_r(\theta)$, which were convolved numerically, for $m_a=1$meV
The Monte Carlo simulation was carried out with $10^9$ events.  The distribution of $\theta$ was determined from sampling $\theta_a$ according to equation (14).  From this, $\theta$ was determined from equation (15).    The theoretical distribution derived from Monte Carlo simulation can be convolved with the point spread function (PSF) of the LAT at 60 MeV, which is approximately 6$^\circ$ for events that convert in front of the tracker.  The angular spread of the theoretical and convolved probability distributions plotted in Figure~\ref{fig:th} may be parametrized by the following quantity:

\begin{equation}\sqrt{\left<\theta^2\right>} = \sqrt{\frac{\sum_i P_i \theta_i^2}{\sum_i P_i}}, \end{equation}
	where $P_i$ is the probability per $i^{th}$ bin, and $\theta_i^2$ is the squared value of $\theta$ in that bin.  We assume, of course, that the distribution is assumed symmetric about 0.  
	
	From inspection of Fig.~\ref{fig:th}, for 1 meV axions, $\sqrt{\left< \theta^2\right>}=1.8^\circ$, and for 10 meV axions, $\sqrt{\left<\theta^2\right>}=2.46^\circ$. Observe that the distribution for the 1 meV axions is diminished, but non-negative, between 2$^\circ$ and 4$^\circ$, due in part to the convolution kernel of the LAT PSF.   

We may determine the $\gamma-$ray energy from the following equation:
\begin{equation} E_\gamma = \frac{1}{2} m\gamma (1+\beta\cos(\theta_a)) \end{equation}

We determine the spectral energy distribution from modifying equation~\ref{eq:sed}, by considering instead of the distance $d$ of the neutron star, the distance $r^\prime$ from the LAT at which the decay vertex $a\to2\gamma$ occurred, which is given by: 

\begin{equation} r^\prime = \sqrt{r^2 + d^2 - 2rd\sin\theta^\prime} \end{equation}

Thus, we obtain:

\begin{equation} E\frac{d\Phi}{dE} = 1.8 \times 10^{-2} \left(\frac{m_a}{\rm eV}\right)^5 \left(\frac{\Delta t}{23.2 \ {\rm s}}\right)\left(\frac{100 \ {\rm pc}}{r^\prime}\right)^2  \left(\frac{2E}{100 \ {\rm MeV}}\right)^4 \left(\frac{S_\sigma(2E)}{10^7 \ {\rm MeV}^2}\right) \ {\rm  cm}^{-2} {\rm s}^{-1}. \label{eq:sed1}\end{equation}

We may note the dependence on the fifth power of the axion mass, as was derived in Ref.~\cite{berenjiAxions}.  The angular probability distribution $P(\theta)$ falls off rapidly with increasing angle.  The smaller the axion mass $m_a$, the narrower the distribution $P(\theta)$ .  It may be observed that the larger the $\gamma$-ray energy, the narrower the angular distribution, as shown in Figure~\ref{fig:th}.  Our limits will provide larger  values of $f_a$ for smaller values of $m_a$ that we choose.  

\section{Projected Limits from the Extended Model for J0108-1431}\label{sec:proj}
\subsection{Simulation Experiment}

We attempt a simulation experiment in order to test the feasibility of determining a signal from a given simulated flux.  One simulation model was considered: the extended model presented earlier.  This simulated model was generated using \emph{gtobssim} from the \emph{ScienceTools}, which was developed using the energy-dependent spatial templates described earlier in this paper in~\ref{sec:extended}, while the spectral model was generated using the function in equation~\ref{eq:sed1}.  The same instrument response functions were used for the data analysis from the experimental observations; In the second case, the extended model for axions was considered.  
   For various injected values of the axion mass $m_a$ into a Monte Carlo simulation, the test statistic ($TS$) and the test statistic for extension ($TS_{ext}$) has been tabulated. The experimental values, for $m_a\simeq 1$ meV, the TS for a point source is 13.56, and the TS$_{ext}=27.11$.   From simulations of the ROI, for a 1 meV axion, the TS=$0.015$, and TS$_{ext}$=0.031.  These values are computed from the likelihood function $\mathcal{L}$, where $\mathcal{L}$ of the likelihood function.   Thus, we establish that the source may be extended .  However, the full simulation of the ROI corresponding to J0108 using $gtobssim$ doesn't match, probably because the fluxes of the point sources are not optimized for this ROI.  
   The value of the test statistic (TS) for extension is for an axion mass of 0.34 meV and flux of 1.84$\cdot 10^{7}$ MeV cm$^{-2}$ s$^{-1}$, where $\mathcal{L}_0$ is the value of the likelihood function, and $\mathcal{L}_{ext}$ is the value of the likelihood function for the extended model.  This value of $TS_{ext}$ is marginally significant, and signals that if extension of such a source exists, then it would be feasible to quantify this numerically.  The TS for detection of an actual source would be 89 for a putative axion mass of 1 meV, which corresponds to a $\sigma=9.4$.   The flux determined from the simulated extended model is 330 $\cdot 10^{-6} $cm$^{-2}$s$^{-1}$MeV$^{-1}$, over a range of gamma-ray energies.   This energy flux compares well with the expected SED in Figure~\ref{fig:sens_10meV}. This mass compares well with the putative axion mass of   0.38 meV derived in the data analysis section.

 \subsection{Simulation of Spectral Model and Spatial Template}     
We chose a near galactic neutron star in order to consider the optimum sensitivity possible with this model.  
The neutron star J0108-1431 was chosen because it is one of the nearest neutron stars at a distance of 130 pc~\cite{xmmJ0108}, and it lies greater than 15$^\circ$ degrees above the galactic plane with celestial coordinates of  ($\alpha,\delta$) = (17.035$^\circ$,-14.351$^\circ$).  This is justified on the grounds that the possible background contamination to the putative signal, the galactic diffuse emission, is greater near the galactic plane.   This neutron star has an age of 1.6$\cdot 10^8$ yr, e.g., Ref.~\cite{TaurisJ0108}.  From the point of view of future Fermi-LAT observations, nearby gamma-ray point sources are less than 1.5$^\circ$ degrees away from J0108-1431~\cite{3FGL} in celestial coordinates.  These features make it possible to consider J0108-1431 as a target for future Fermi constraints on axions.

\begin{figure}
\begin{centering}
\includegraphics[width=10cm]{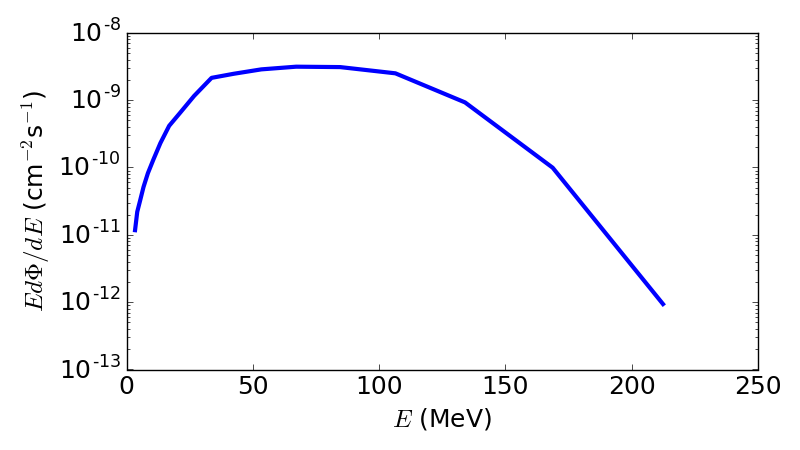}
	\caption{PSF-cls onvolved spectrum for extended emission of 10 meV axions from neutron star J0108-1431, see text for details.}\label{fig:spectrum}
\end{centering}
\end{figure}

The spectral model corresponding to equation~\ref{eq:sed1} is plotted in Figure~\ref{fig:spectrum}.  Gamma-rays with energy 60-200 MeV are produced from this model.  The spectral model peaks around 75 MeV. The spectral model is convolved with the point spread function.

\begin{figure}
\begin{centering}
	\includegraphics[width=3cm]{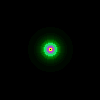}
	\includegraphics[width=3cm]{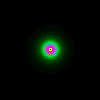}
	\includegraphics[width=3cm]{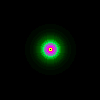}
	\includegraphics[width=3cm]{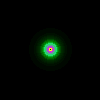}
	\includegraphics[width=3cm]{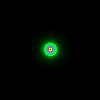}
	\includegraphics[width=0.2cm]{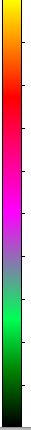}
	\caption{Spatial distribution maps of gamma-rays around neutron star J0108-1431. $E=$36.27, 53.00, 77.46, 113.20, 165.44.   The units of the colorbar are density (normalized to 1) on a logarithmic scale. The pixels are 0.4$^\circ\times0.4^\circ$.}\label{fig:mapcube}
\end{centering}
\end{figure}

We generate a spatial template spatial map at the following 7 log-spaced gamma ray energies in units of MeV: 24.81, 36.27, 53.00, 77.46, 113.20, 165.44, 241.78.  We generate a spatial template of 100 pixels x 100 pixels, with a bin size of 0.4$^\circ$/pixel.  The map-cube function corresponding to neutron star J0108-1431 is shown in Figure~\ref{fig:mapcube}.  

Observe that the spatial distribution around the images in Fig.~\ref{fig:mapcube} have energy-dependent radii.  It is all the more important to consider this when optimizing an astrophysical analysis.   

\begin{figure}
\begin{centering}
\includegraphics[width=4in]{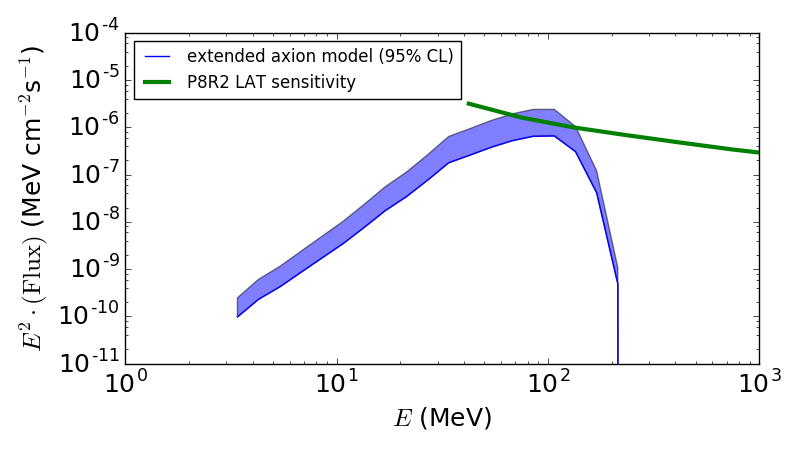}
\caption{The expected spectral energy distribution of 10 meV axions from the spatially-extended model compared to the point source sensitivity of the LAT.  The region is determined by 10\% hardening or softening of the spectrum curve.  The LAT curve is the 10-year sensitivity using Fermi-LAT Pass 8 instrument response functions, with a region corresponding to 95\% CL.}\label{fig:sens_10meV}
\end{centering}
\end{figure}

Henceforth in this article, all analyses performed with Fermi LAT data was done with the ScienceTools-v10r0p5-fssc-20150518A-source, Ubuntu version, as downloaded from NASA's FSSC website~\cite{ScienceTools}.  
In Figure~\ref{fig:sens_10meV} we plot the expected sensitivity of the model.  With the projected 10-year sensitivity of Fermi LAT using Pass 8 instrument response functions, it may be possible in principle to observe or set upper limits on the axion mass below 10 meV.  
	 
\section{Observations\label{sec:obs}}
\subsection{Methods}
We have analyzed the LAT observations of neutron star J0108-1431, at a distance of 240 pc, with
The neutron star J0108-1431 was chosen for analysis because nearby point sources less than 1.5$^\circ$ degrees away, it is one of the nearest neutron stars 130 pc~\cite{xmmJ0108}, and it lies greater than 15$^\circ$ degrees above the galactic plane. 

We use a sample of events spanning the interval 239557417-489220014 MET, for a duration of 150 Ms or 7.9 yr, using $\gamma$-rays with energies between 30 MeV and 292 MeV. We use the ScienceTools~\cite{ScienceTools} in the analysis of the ROI.   We select front-converting events, which have an improved PSF profile over back-converting events.  We perform a binned analysis, with 6 bins in the selected energy range.  We choose a spatial binning of 0.8$^\circ$/pixel for the analysis.  We use the Pass8 instrument response function \emph{P8R2\_SOURCE\_V6}.  We found that using PSF event class types did not prove to optimize the analysis,  since front-converting events are predominant in this energy range, and therefore we chose the \emph{P8R2\_SOURCE\_V6} with front events.  We consider energy dispersion in the fit for the putative extended source modeled according to the model presented in Section~\ref{sec:extended}.  We include in the analysis the point-like sources included in the Third Fermi LAT catalog (3FGL)~\cite{3FGL}.  The 3FGL point sources were modeled with energy dispersion.  The 3FGL model was derived in a fit above 100 MeV~\cite{3FGL}.   We do not consider energy dispersion for the galactic diffuse and isotropic diffuse models (when using the fitted parameters), since they already have energy dispersion factored in.   All of the 3FGL point sources have fixed spectral indices.   Those 3FGL sources within 15$^\circ$ of J0108-1431 have free normalizations, otherwise the normalizations are fixed.  
   
The log-likelihood test statistic is used:
\begin{equation} TS = -2\log\left(L_0/L_1\right), \end{equation}
	where $L_0$ is the likelihood without the putative neutron star axion source, and $L_1$ is the likelihood with the the putative neutron star axion source.  
The binned likelihood function can be expressed as:
\begin{equation} \mathcal{L} = \prod_i \frac{M_i^{n_i}e^{-M_i}}{n_i!} \end{equation}
where $M_i$ are the predicted counts in each bin, and n i are the detected counts in
each bin. This is essentially of product of probabilities, based upon the assumption
that the observed number of counts in each bin may be characterized by a Poisson
distribution. This holds in the case of a small number of counts. In the case of
unbinned likelihood, the bin sizes get infinitesimally small, such that $n_i$ is either 0 or 1.

We use the galactic diffuse model \emph{gll\_iem\_v06.fits} and the isotropic diffuse model \\ {\itshape iso\_p8r2\_source\_v6\_front\_v06.txt}, described in Ref.~\cite{diffuseModels}. The former extends down to energies of 58 MeV, and the latter extends down to 36 MeV.  These diffuse models have been extrapolated as a Heaviside function down to 30.0 MeV in order to enable this analysis.  Al though a power-law extrapolation might be more accurate, the spectral index was very small in any case, and thus no significant error was introduced by the Heaviside approximation for the diffuse.  


We iteratively fit the spectral parameters for the ROI in three steps.   In the first step, all 3FGL point sources within 15$^\circ$ of the ROI center are fit, with only normalizations varying, along with the diffuse models and the extended axion model, using the DRMNFB minimizer~\cite{DRMNFB}.  In the second step, 3FGL point sources further than 15$^\circ$ from the ROI center are fit, with only normalizations varying, in addition to the diffuse models and the extended axion model, using the DRMNFB minimizer. In the third step, the 3FGL point sources within 10$^\circ$ of the ROI center are fit, with free spectral indices as well as normalizations, in addition to the diffuse and extended models, using the MINUIT minimizer~\cite{MINUIT}.  The statistical uncertainties are below 10\% within the fit energy range, as shown in Figure~\ref{fig:spectralResiduals}.   
\subsection{Results}\label{sec:results}
A search is made for the axion signal in gamma-rays in the specified energy range in a ROI corresponding to J0108-1431.  Within statistical and systematic uncertainties, no signal is detected for this ROI, therefore, we proceed to set limits on the axion mass. 
  
\begin{figure}

	\begin{centering}
	\includegraphics[width=7.5cm]{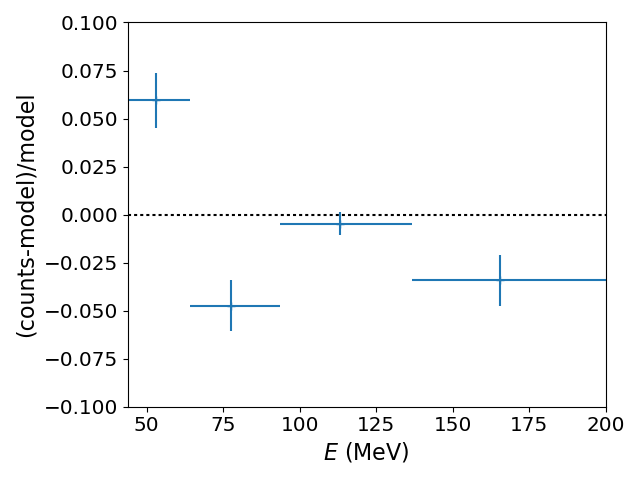}
	\caption{Spectral residuals in the $40^\circ\times40^\circ$ ROI centered in J0108-1431.  Error bars include statistical and systematic sources of error.  }\label{fig:spectralResiduals}
	\end{centering}
\end{figure}

\begin{figure}
	\begin{centering}
		\includegraphics[width=7.5cm]{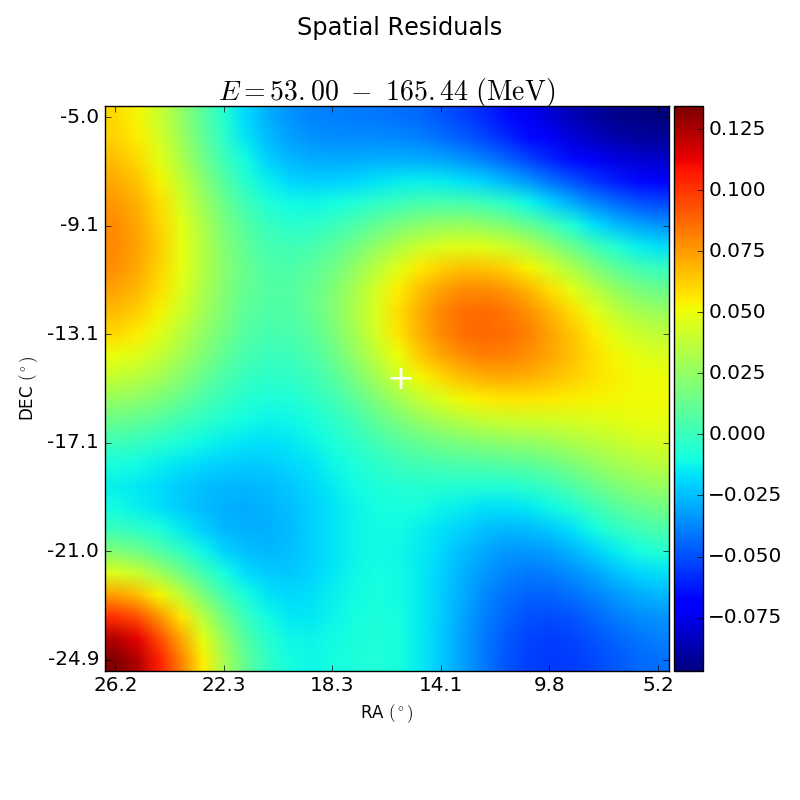}
		\caption{Spatial residuals over the energy range $E (MeV)=53.0-165.4$, in the 20$^\circ$ $\times$ 20$^\circ$ ROI centered at the position of J0108-1421, convolved with a Gaussian smoothing kernel of $\sigma=3.2^\circ$.  The $z-$axis scale is the fraction (counts-model)/model}\label{fig:spatialResiduals} 
	\end{centering}
\end{figure}

The spectral residuals, as shown in Figure~\ref{fig:spectralResiduals}, are within 5\% over the energy range 43.4 MeV to 200 MeV.  The spatial residuals (residual map), as shown in Figure~\ref{fig:spatialResiduals}, has values in each pixel which are within $\pm$ 15\% of 0, the latter representing perfect agreement between spatial model counts and measured count distributions.  The analysis-derived upper limits on the flux and the mass are given in Table~\ref{tab:limits}. 

\begin{table}
	\begin{centering}
\begin{tabular}{|l|l|}
\hline
	95\% CL value \\
\hline
$\Phi$ & $>$ 3.5$\times 10^{-12}$ cm$^{-2}$s$^{-1}$\\
\hline
$m_a$ & $<$ 0.76 meV \\
	$g_{ann}$ & $>$ $7.6\times 10^{-12}$ \\
\hline
\end{tabular}
\caption{For neutron star J0108-1431, the flux upper limit and the upper limit on the axion mass.}\label{tab:limits}
\end{centering}
\end{table}

The Monte Carlo distribution of photons is binned in angle $\theta$ and energy$E$.  It is the angular distribution which is radially symmetric.  One may consider variations in the
model parameters for the Monte Carlo derived energy and angular distribution.  Qualitatively, model parameters such as
 the 3FGL source may be examined as well.  
 \begin{itemize}
                \item There may be possible degeneracy between fitting the diffuse sources
 and the extended source template angular distribution.
                \item Due to a large number of sources within the ROI, there are a large number of free parameters, and the Minuit2 fit may not have found the true minimum of the likelihood function.
        \end{itemize} 
	\section{Discussion}\label{sec:disc}
\begin{figure}
\begin{centering}
\includegraphics[width=4in]{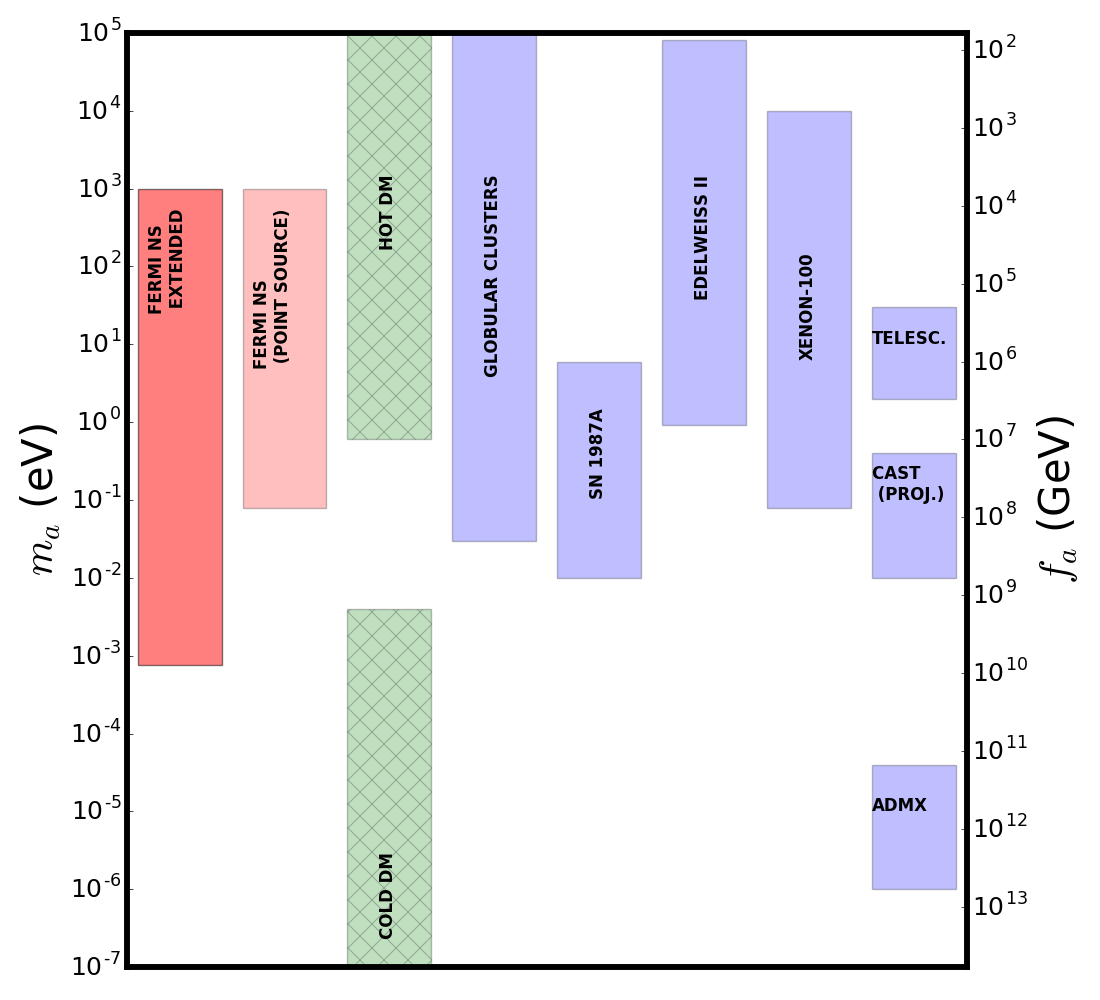}
\caption{Comparison of exclusion ranges compared with the possible range of masses presented in this article. Exclusion regions for axions: the Fermi-LAT point source limits (light red) from neutron stars, projected limits from the Fermi LAT using the spatially-extended model (dark red) of the neutron star J0108-1431, compared with previous astrophysical limits.}\label{fig:limits}
\end{centering}
\end{figure}

	The model-dependent observational limits derived here for the hypothesis of extension of $m_a<0.76$ meV at the 95\% C.L. are a substantial improvement upon the previous Fermi-LAT point-source limit with neutron stars~\cite{berenjiAxions}.  The point-source limits for this source under the same data analysis conditions as for the extended source model would not be changed much from the previous point-source limits, as the energies where the spectrum has a considerable contribution (i.e., above 60 MeV) were considered in the previous analysis.  Furthermore, the data-dependent limits also fall below the projected limits derived here.    From a consideration of Figure~\ref{fig:limits}, the limits from the Fermi LAT using 7.9 years of observations are improved by a factor of nearly 100 with respect to the point source limits.  Also, these limits fall in a range that has not yet been excluded by previous observations, and represent a substantial improvement over the SN 1987A-derived upper limit of $m_a<16$ meV~\cite{SNbounds}.  If the hypothesis of extension is valid, however low the flux may be, then we report a highly significant detection.  This should be explored in a possible followup study.  The signal for J0108 could be contaminated with known nearby point sources according to the 3FGL Fermi-LAT catalog.  It should be noted that the possible detection of a signal depends on the extended model energy-dependent spatial templates shown and derived earlier in this study, and that the point source study of a putative signal are probably to na\"ive  for searching for axions.  Needless to say, future studies should focus on neutron stars with farther distances and that are also younger in age.  
	
	It deserves to be mentioned that assuming somewhat higher or lower temperatures could alter this limit somewhat.  The simulation-derived limit of 10 meV is not at the minimum range possible, according to this model, but suggests an upper bound on values that could be derived in principle. 

Interestingly, the limit constrains the allowed parameter space for axions as cold dark matter.  These projected limits are better than the bounds of $m_a<20$ meV reported by CAST~\cite{CAST2015}.  While the ADMX projection excludes a smaller range of masses, it is probing DFSZ axions not KSVZ axions.  
This model sets more restrictive limits than this and other neutron star cooling observations~\cite{sedrakian}.   Neutron star cooling by axions is a quite distinct process than the process of emitted axions decaying outside the neutron star, for which current limits have been reported as $m_a<60$ meV.  
In future observations, the projected limits could potentially be improved by statistically combining limits from multiple neutron stars, as shown in Ref.~\cite{berenjiAxions}.  
	Although there may be some uncertainty over the precise temperature of the neutron star, which we assume as $T=20$ MeV, this applies generally to neutron stars with hadronic physics.  In extended models of neutron stars, which contain free quarks in a QCD phase~\cite{Alford2009}, it is generally accepted that there is a range of temperatures which are higher, generally between 10 MeV to 60 MeV~\cite{ruster2005phase}
	In this article, we set limits on $f_a\sim 10^{10}$ GeV.  Axions with masses $\sim1$meV may be a source of dark matter, although it cannot comprise the majority of the dark matter because the abundance is too low~\cite{graham2013}.  This bound is much stronger than the weak upper bound of $f_a\lesssim 10^{12}$ GeV from cosmological arguments~\cite{dine}.
	In future work, a consideration of the possible background signal from the pulsar in addition to the axion signal could result in a better limit on the axion mass.  In addition, simulation studies could enhance the selection of neutron star targets that would yield the best limits.  If equation~\ref{mafa} is relaxed, so that both $m_a$ and $f_a$ are free parameters, then ALPs may also be considered in a generalization of the model presented here.  
We quantitatively consider the effect of variations in these parameters on the axion limits in Section~\ref{sec:obs}. 
Qualitatively, increasing (decreasing) the assumed $T$ would tend to shift the spectrum towards higher (lower) energies.  The model flux depends on $\omega^4 S_\sigma(\omega;\mu,T)$, which increases with $T$, but the timescale depends on $\left(\int d\omega \ \omega S_\sigma(\omega;\mu,T)\right)^{-1}$, which decreases with $T$.  Thus, a simple calculation finds that the limits on $m_a$ would be smaller for $T = 50$ MeV, and larger for $T=10$ MeV.  The order of magnitude of the limits would still be the same for these changes in temperature.  If the assumed $T$ were higher, the \emph{Fermi} LAT could be used to set more stringent limits.   For $T=10$ MeV, we estimate the limits would be a factor of 4 less restrictive.  
Increasing the degeneracy parameter $\mu$ would tend to decrease the amplitude of the spin-structure function.  At $\mu/T = 11$, the limits would be larger, and at $\mu/T = 9$, the limits would be smaller.  Changing the $k$ parameter would not affect the limits substantially.

	Similar limits have been achieved by other recent axion searches. Supernova constraints for SN 1987A yield a limit of $m_a \lesssim 16$ meV ~\cite{Raffelt1987A}.   From vacuum  magnetic birefringence, a limit of $m_a<5\times 10^{-4}$ eV has been reached~\cite{birefringence}.  From magnetically induced dichroism, $m_a=1-1.5$ meV has been reached~\cite{zavattini}.  From globular cluster studies, a limit of  $m_a \sin^2\beta < 15$ meV has been attained~\cite{viaux}.  From the white dwarf luminosity function, an inclusion range of $2.5\lesssim m_a \sin^2 \beta \lesssim 7.5$ meV has been achieved~\cite{WD}\cite{WDisern}.   From the supernova remnant Cas A, the axion mass was set at $m_a = 2.3\pm 0.4$ meV$/C_n$~\cite{CasA}. These limits are complementary to the result shown in our work, and we hope that better limits could be obtained in future work.   

\pagebreak

\section{Acknowledgments}
The \textit{Fermi}-LAT Collaboration acknowledges support for LAT development, operation and data analysis from NASA and DOE (United States), CEA/Irfu and IN2P3/CNRS (France), ASI and INFN (Italy), MEXT, KEK, and JAXA (Japan), and the K.A.~Wallenberg Foundation, the Swedish Research Council and the National Space Board (Sweden). Science analysis support in the operations phase from INAF (Italy) and CNES (France) is also gratefully acknowledged. This work performed in part under DOE Contract DE-AC02-76SF00515.
BB is thankful for constructive comments from Seth Digel and Eric Charles of SLAC National Accelerator Laboratory.  I acknowledge the conveners at CERN for an invited talk during May 2017. BB acknowledges the support to fiziSim LLC in a generous grant from Orthopedic and Arthritic Center, Oxnard, California.  BB acknowledges support from California State University, Los Angeles Department of Physics and Astronomy in the College of Natural and Social Sciences.   



\begin{thebibliography}{50}
		\bibliography{refs1.bib}
		\end{thebibliography}

\begin{thebibliography}{10}

\bibitem{PQ}
R.~Peccei and H.~Quinn,
\newblock Phys. Rev. Lett. {\bfseries 38}, 1440 (1977).

\bibitem{Sikivie201122}
P.~Sikivie,
\newblock Physics Letters B {\bfseries 695}, 22  (2011).

\bibitem{Skobolev2000}
V.~V. {Skobelev},
\newblock Physics of Atomic Nuclei {\bfseries 63}, 1963 (2000).

\bibitem{berenjiAxions}
B.~Berenji, J.~Gaskins, and M.~Meyer,
\newblock Physical Review D {\bfseries 93}, 045019 (2016).

\bibitem{instrumentPaper}
W.~B. Atwood {\em et~al.},
\newblock The Astrophysical Journal {\bfseries 697}, 1071 (2009).

\bibitem{lande2012search}
J.~Lande {\em et~al.},
\newblock The Astrophysical Journal {\bfseries 756}, 5 (2012).

\bibitem{ext2017}
M.~Ackermann {\em et~al.},
\newblock ApJ {\bfseries 843} (2017).

\bibitem{extHiLat}
R.~Caputo, M.~Meyer, M.~Wood, and J.~Biteau,
\newblock The Astrophysical Journal Supplement Series {\bfseries 237} (2018).

\bibitem{AbazajianDM}
K.~N. Abazajian and M.~Kaplinghat,
\newblock Phys. Rev. D {\bfseries 86}, 083511 (2012).

\bibitem{andromedaExtended}
M.~Ackermann {\em et~al.},
\newblock The Astrophysical Journal {\bfseries 836}, 208 (2017).

\bibitem{giannotti}
M.~Giannotti, L.~Duffy, and R.~Nita,
\newblock Journal of Cosmology and Astroparticle Physics {\bfseries 2011}, 015
  (2011).

\bibitem{meyer2013first}
M.~Meyer, D.~Horns, and M.~Raue,
\newblock Physical Review D {\bfseries 87}, 035027 (2013).

\bibitem{raffeltConversion}
G.~Raffelt and L.~Stodolsky,
\newblock Physical Review D {\bfseries 37} (1988).

\bibitem{raffelt}
G.~G. Raffelt,
\newblock {\itshape Stars as Laboratories for Fundamental Physics: The
  Astrophysics of Neutrinos, Axions, and Other Weakly Interacting Particles}
  (University of Chicago Press, 1996).

\bibitem{raffelt2011meV}
G.~G. Raffelt, J.~Redondo, and N.~V. Maira,
\newblock Physical Review D {\bfseries 84}, 103008 (2011).

\bibitem{redondo2012journey}
J.~Redondo, G.~Raffelt, and N.~V. Maira,
\newblock Journey at the axion mev mass frontier,
\newblock in {\itshape Journal of Physics: Conference Series} Vol. 375, p.
  022004, IOP Publishing, 2012.

\bibitem{sedrakian}
A.~Sedrakian,
\newblock Phys. Rev. D {\bfseries 93}, 065044 (2016).

\bibitem{pass8}
W.~Atwood {\em et~al.},
\newblock 4$^{th}$ Fermi Symposium {\bfseries eConf C121028} (2012),
  astro-ph/1303.3514.

\bibitem{lloyd}
S.~J. Lloyd, P.~M. Chadwick, and A.~M. Brown,
\newblock Physical Review D {\bfseries 100} (2019).

\bibitem{hamaguchi}
K.~Hamaguchi, N.~Nagata, K.~Yanagi, and J.~Zheng,
\newblock Physical Review D {\bfseries 98} (2018).

\bibitem{abdelhameed}
A.~H. Abdelhameed {\em et~al.},
\newblock The European Physical Journal C {\bfseries 80} (2020).

\bibitem{ksvz1}
J.~E. Kim,
\newblock Phys. Rev. Lett. {\bfseries 43}, 103 (1979).

\bibitem{ksvz2}
M.~A. Shifman, A.~I. Vainstein, and V.~I. Zakharov,
\newblock Nuclear Physics {\bfseries B1966} (1980).

\bibitem{dfsz1}
M.~Dine, W.~Fischler, and M.~Srednicki,
\newblock Physics Letters B {\bfseries 104}, 199 (1981).

\bibitem{dfsz2}
A.~Zhitnitsky,
\newblock Sov.J.Nucl.Phys. {\bfseries 31}, 260 (1980).

\bibitem{RPP}
M.~Tanabashi and others (Particle Data~Group),
\newblock Phys. Rev. D {\bfseries 98} (2018).

\bibitem{nambu}
Y.~Nambu and M.~Sasaki,
\newblock Phys. Rev. D {\bfseries 44} (1991).

\bibitem{sanchez2009hints}
M.~S{\'a}nchez-Conde, D.~Paneque, E.~Bloom, F.~Prada, and A.~Dominguez,
\newblock Physical Review D {\bfseries 79}, 123511 (2009).

\bibitem{SNbounds}
G.~G. Raffelt,
\newblock Lect. Notes Phys. {\bfseries 741}, 51–71 (2008), hep-ph/0611350.

\bibitem{uw}
C.~Hanhart, D.~R. Phillips, and S.~Reddy,
\newblock Physics Letters B {\bfseries 499}, 9  (2001), astro-ph/0003445.

\bibitem{axionBook}
G.~G. Raffelt,
\newblock {\itshape Astrophysical Methods to Constrain Axions and Other Novel
  Particle Phenomena} (North--Holland, 1990).

\bibitem{ruster2005phase}
S.~B. R{\"u}ster, V.~Werth, M.~Buballa, I.~A. Shovkovy, and D.~H. Rischke,
\newblock Physical Review D {\bfseries 72}, 034004 (2005), hep-ph/0503184.

\bibitem{pavlov2009}
G.~Pavlov, O.~Kargaltsev, J.~Wong, and G.~Garmire,
\newblock The Astrophysical Journal {\bfseries 691}, 458 (2009).

\bibitem{garmire}
G.~P. Garmire,
\newblock ApJ {\bfseries 691}, 458–464 (2009).

\bibitem{riper}
K.~A. {van Riper}, B.~{Link}, and R.~I. {Epstein},
\newblock Astrophys. J. {\bfseries 448}, 294 (1995), astro-ph/9404060.

\bibitem{pionDecay}
K.~T. McDonald, 2019.

\bibitem{xmmJ0108}
B.~Posselt {\em et~al.},
\newblock The Astrophysical Journal {\bfseries 761}, 117 (2012).

\bibitem{TaurisJ0108}
T.~M. {Tauris} {\em et~al.},
\newblock Ap.J.L  (1994).

\bibitem{3FGL}
F.~Acero {\em et~al.},
\newblock The Astrophysical Journal Supplement Series {\bfseries 218}, 23
  (2015).

\bibitem{ScienceTools}
Fssc: Fermi data : Online documentation : Science tools: Cicerone, 2015.

\bibitem{diffuseModels}
{LAT Background Models}, 2016.

\bibitem{DRMNFB}
Likelihood model fitting, 2011.

\bibitem{MINUIT}
F.~James and M.~Winkler,
\newblock Minuit user's guide, 2011.

\bibitem{CAST2015}
CAST Collaboration, M.~Arik {\em et~al.},
\newblock Phys. Rev. D {\bfseries 92}, 021101 (2015).

\bibitem{Alford2009}
M.~G. Alford,
\newblock Nuclear Physics A {\bfseries 830}, 385c  (2009),
\newblock Quark Matter 2009.

\bibitem{graham2013}
P.~W. Graham and S.~Rajendran,
\newblock Phys. Rev. D {\bfseries 88}, 035023 (2013).

\bibitem{dine}
M.~Dine and W.~Fichsler,
\newblock Phys. Lett. {\bfseries B120}, 127 (1983).

\bibitem{Raffelt1987A}
G.~Raffelt,
\newblock Lect. Notes Phys. {\bfseries 741} (2008).

\bibitem{birefringence}
L.~Maiani {\em et~al.},
\newblock Phys. Lett. {\bfseries B175} (1986).

\bibitem{zavattini}
E.~Zavattini {\em et~al.},
\newblock Phys. Rev. Lett. {\bfseries 96} (2006).

\bibitem{viaux}
N.~Viaux {\em et~al.},
\newblock Phys. Rev. Lett. {\bfseries 111} (2013).

\bibitem{WD}
M.~M. Bertolami, 2014.

\bibitem{WDisern}
J.~Isern {\em et~al.}, 2008.

\bibitem{CasA}
L.~Leinson,
\newblock JCAP {\bfseries 1408} (2014).

\end{thebibliography}

\end{document}